\documentclass[prd,aps,preprintnumbers,floats,floatfix,superscriptaddress,preprintnumbers,
showpacs,eqsecnum,nofootinbib,twocolumn]{revtex4-1}
\usepackage[utf8]{inputenc}
\usepackage{latexsym,array,theorem,mathrsfs,bm,float}
\usepackage{psfrag}
\usepackage{amsfonts,amsmath,amssymb,latexsym,array,afterpage,theorem,color,graphicx,mathrsfs,bm,float,tabularx,here,
enumerate,multirow,comment}

\newcommand{\f}[2]{\frac{#1}{#2}}

\renewcommand{\)}{\right)}

\renewcommand{\O}{\Omega}

\renewcommand{\t}{\tau}

\renewcommand{\r}{\rho}

\begin{document}
\baselineskip=12pt

\preprint{OCU-PHYS-518}
\preprint{AP-GR-160}
\preprint{NITEP 64}
\date{\today}

%%%%%%%%%%%%%%%%%%%%%%%
%<<<<<<<<<<<<< TITLE >>>>>>>>>>>>>>>%
%%%%%%%%%%%%%%%%%%%%%%%
\title{
Do black hole shadows merge?
}
%%%%%%%%%%%%%%%%%%%%%%%
%<<<<<<<<<<< AUTHOR >>>>>>>>>>>>>%
%%%%%%%%%%%%%%%%%%%%%%%
\author{Kazumasa Okabayashi}
\email{okabayashi"at"ka.osaka-cu.ac.jp}
\author{Nobuyuki Asaka}
\email{asaka"at"sci.osaka-cu.ac.jp}
\affiliation{
Department of Mathematics and Physics, Graduate School of Science, Osaka City University, Sumiyoshi, Osaka City 558-8585, Japan}
\author{Ken-ichi Nakao}
\email{knakao"at"sci.osaka-cu.ac.jp}
\affiliation{
Department of Mathematics and Physics, Graduate School of Science, Osaka City University, Sumiyoshi, Osaka City 558-8585, Japan}
\affiliation{
Nambu Yoichiro Institute of Theoretical and Experimental Physics, Osaka City University, Sumiyoshi, Osaka City 558-8585, Japan}
%<<<<<<<<<<<<< DATE >>>>>>>>>>>>>>>%
\date{\today}
%%%%%%%%%%%%%%%%%%%%%%%%
%<<<<<<<<<<< ABSTRACT >>>>>>>>>>>>>%
%%%%%%%%%%%%%%%%%%%%%%%%
\begin{abstract}
The so-called black hole shadow is not a silhouette of a black hole but an image of a collapsing object or a white hole. Hence
it is nontrivial whether black hole shadows merge with each other when black holes coalesce with each other.
In this paper, by analyzing the null geodesic generators of the event horizon in Kastor-Traschen spacetime which describes a coalescence of black boles,
we see that observers who will never see a merger of black hole shadows exist.
\end{abstract}

\maketitle

%\tableofcontents
%\setcounter{page}{1}
%\newpage

%%%%%%%%%%%%%%%%%%%%%%%%%%%%%%%%%%%%%%%%%%%%%%%%%%%%%%%%%%%%%%%%%%%%%%%%%
\section{INTRODUCTION}
\label{introduction}
%%%%%%%%%%%%%%%%%%%%%%%%%%%%%%%%%%%%%%%%%%%%%%%%%%%%%%%%%%%%%%%%%%%%%%%%%
The black hole is one of the most fascinating predictions of general relativity. It is defined as a complement of the causal past of future null infinity.
This definition implies that any observers outside a black hole cannot receive any physical influences caused in the black hole.
General relativity predicts that black holes will form through gravitational collapse of massive objects, but
in view of any observers outside black holes,
those massive objects will continue to collapse forever and no black hole forms.
No black hole has ever formed in our view, although there will be many black holes in our Universe.

The so-called black hole shadow
taken by Event Horizon Telescope Collaboration \cite{collaboration2019m87} is , exactly speaking, not a silhouette of a black hole but will be
unresolved images of collapsing objects or
possibly an image of a black hole mimicker. (In this paper, hereafter, we will not consider the possibility of black hole mimickers in order to make discussions simple.)
What we have ever observed will be collapsing objects and their neighboring exterior domains.
By virtue of the no-hair nature of black holes, the spacetime geometry of the neighboring exterior domains
of collapsing objects will asymptotically approach Kerr-Newman family. Even before the collapsing objects form black holes, we will observe the same phenomena
as those occurring around black holes. However, we cannot rule out the possibility that the collapsing object will stop collapsing after we stop observing it \cite{PhysRevD.99.044027}.
If it is a black hole, we cannot confirm the fact that it is a black hole by its definition. On the other hand, a black hole model to explain observational data is, in principle, falsifiable. The black hole model can be scientific in Popper's sense.

In theoretical calculation, a black hole shadow is defined as an image made of null geodesics emanating from events on a sphere with the horizon radius
(see Ref.~\cite{PhysRevD.99.044027} and also Appendix \ref{BH-shadow}).
Here note that the sphere with the horizon radius is not an event horizon but a white hole horizon, since there is no future-directed null geodesic
emanating from the event horizon to observers outside the black hole. Following this definition, the black hole shadows in multiblack hole systems
have been theoretically studied in the case for the static solution in \cite{Yumoto_2012}, the stationary solution in \cite{PhysRevD.98.044053},
and the colliding black holes in \cite{PhysRevD.84.063008}. In static or stationary cases, it is known that black hole shadows do not merge with each other. By contrast, in the case
for colliding black holes, it has been claimed that the shadows collide with each other. However, since black hole shadows found in our Universe will be images of
collapsing objects and a black hole shadow obtained by theoretical calculation is an image of a white hole, it is a nontrivial question whether
black hole shadows merge with each other even if black holes will merge.

There are several exact solutions describing colliding black holes \cite{PhysRevD.47.5370,PhysRevLett.104.131101, Behrndt_2003}.
Among these solutions, Kastor-Traschen (KT) solution is one of the simplest solution describing colliding black holes \cite{PhysRevD.47.5370} .
The global structure of the KT solution was studied in
\cite{PhysRevD.49.840,PhysRevD.52.796,PhysRevLett.74.630,PhysRevD.58.121501}
but
has not been seriously considered from an observational point of view. We revisit this issue and reconsider what the global structure implies observationally.

In this paper, we focus on the KT solution with two ``particles" with identical mass in order to show that black hole shadows do not necessarily merge.
The behavior of black hole shadows in this system was numerically analyzed in detail by Yumoto \textit{et.al.} \cite{Yumoto_2012},
and their results, especially Fig.~9 in their paper, indicate that black hole shadows
do not merge, but they expect that the black hole shadows eventually merge with each other. In this paper, we analyze behavior of null geodesic generators of the event horizon
and would like to claim opposite of their expectation, i.e., that there are observers who will not see the coalescence of black hole shadows even if the black holes will coalesce into
one.  In Sec.~\ref{Kastor-Traschen_spacetime}, we briefly review the KT solution. In Sec.~\ref{RNdS}, the case of one particle is reviewed
since it is useful to understand the global structure in the case of two particles of our interest.
Then, in Sec.~\ref{2-particles}, we study the case of two particles, especially the global structure of a timelike hypersurface that extends through just the middle of the two particles.
Finally, Sec.\ref{concluding_remarks} is devoted to concluding remarks. Throughout this paper,
we use the geometrical units of $c=G=1$ and follow \cite{misner1971gravitation}
for the notations.

%%%%%%%%%%%%%%%%%%%%%%%%%%%%%%%%%%%%%%%%%%%%%%%%%%%%%%%%%%%%%%%%%%%%%%%%%
\section{Kastor-Traschen solution}
\label{Kastor-Traschen_spacetime}
%%%%%%%%%%%%%%%%%%%%%%%%%%%%%%%%%%%%%%%%%%%%%%%%%%%%%%%%%%%%%%%%%%%%%%%%%
The KT solution, equivaleently, the KT spacetime is an exact solution of the Einstein-Maxwell system with positive cosmological constant $\Lambda$,
which describes the motion of extremely charged ``particles".
$M_i$ denotes the mass of the $i$th particle located at $(x,y,z)=(x_i,y_i,z_i)$ in comoving cosmological Cartesian coordinates,
and its electric charge is equal to $M_i$. In this paper, we assume $M_i>0$.
There are two classes: one is represented in the contracting cosmological time coordinate $\t_-$, and the other is given in the expanding cosmological time coordinate $\t_+$.
The metric and the gauge field $A_\mu$ are given in the form
\begin{align}
 ds^2=-U^{-2}d \t_{\pm}^2+U^2 \left( dx^2+dy^2+dz^2 \right),
	\label{KTtaumet}
\end{align}
and
\begin{align}
A_\mu=\left(U^{-1},0,0,0\right),
\end{align}
where $H=\sqrt{\Lambda/3}$, and
\begin{align}
 U=\pm H\t_\pm+\sum_{i=1}^N \f{M_{i}}{r_{i}}
\end{align}
with
\begin{align}
 r_{i}=\sqrt{\left(x-x_{i}\right)^2+\left(y-y_{i} \right)^2+\left(z-z_{i} \right)^2}.
\end{align}
It is a remarkable property of the multiparticle solution with the contracting time coordinate that if the total mass is less than $1/4H$,
particles are black holes and will coalesce with each other to form one black hole.

Kretschmann curvature invariant diverges at $U=0$ in the domain of $|r_i|<\infty$
and that means $U=0$ gives curvature singularity in this domain\cite{PhysRevD.52.796}.
Hence, the covered domains by $\t_\pm$ are restricted by the condition $U>0$ which leads to
\begin{align}
-\infty < \t_- <\sum_{i=1}^N\frac{M_i}{Hr_i}, \\
-\sum_{i=1}^N\frac{M_i}{Hr_i}<\t_+<\infty.
\end{align}

%%%%%%%%%%%%%%%%%%%%%%%%%%%%%%%%%%%%%%%%%%%%%%%%%%%%%%%%%%%%%%%%%%%%%%%%%
\section{Case of one particle}
\label{RNdS}
%%%%%%%%%%%%%%%%%%%%%%%%%%%%%%%%%%%%%%%%%%%%%%%%%%%%%%%%%%%%%%%%%%%%%%%%%
The KT spacetime with one particle is equivalent to the extreme Reissner-Norstr\"{o}m-de Sitter (RNdS) spacetime
which has been well studied by Brill and Hayward \cite{Brill_1994}. Since the RNdS spacetime is useful in understanding the global structure of multiparticle cases,
we will review it in a bit detail here.

%%%%%%%%%%%
\begin{figure}[htbp]
\includegraphics[width=8cm,height=8cm,keepaspectratio]{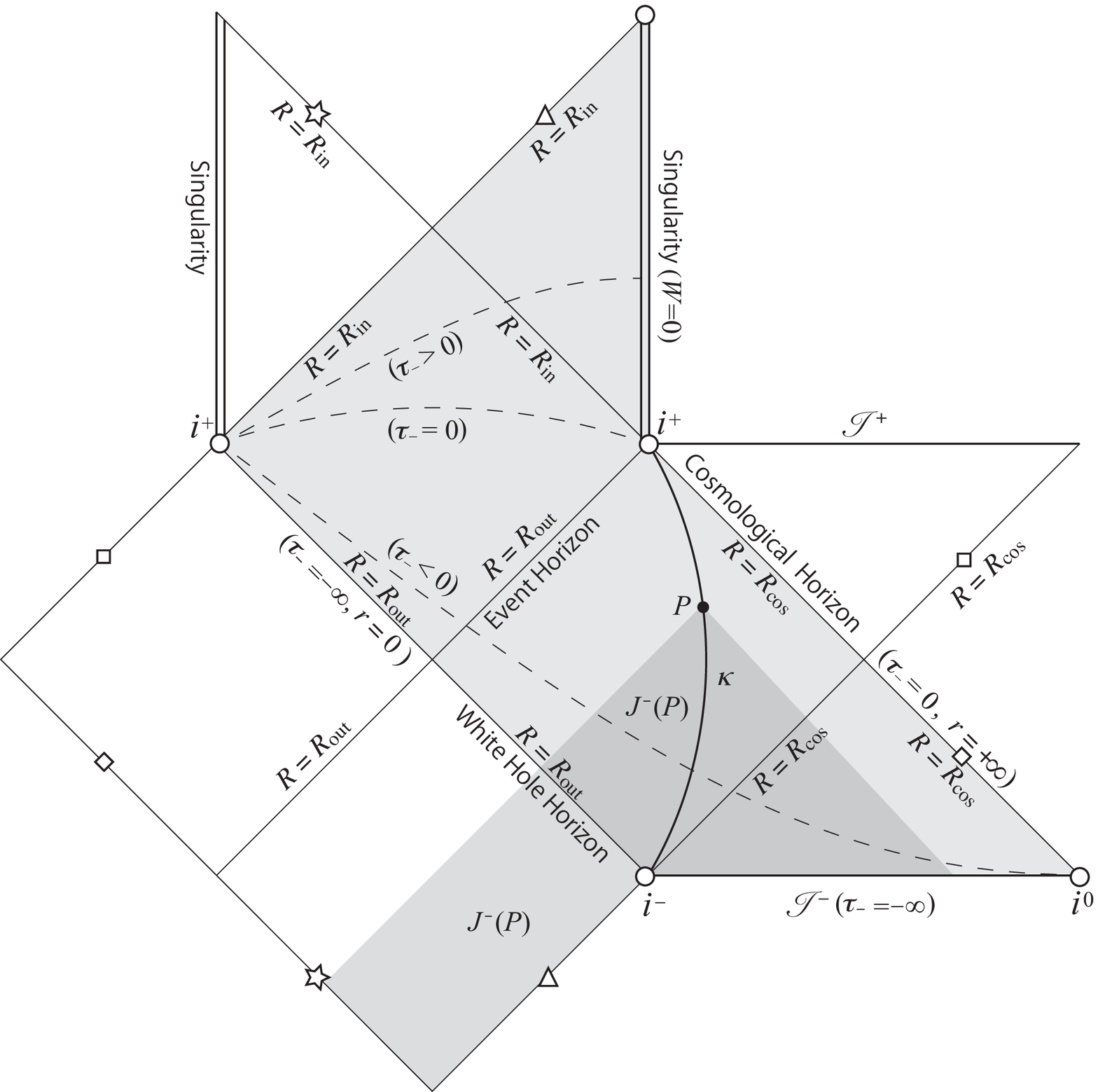}
\caption{The conformal diagram of the maximally extended RNdS spacetime is depicted. Each point in this diagram is a sphere of constant $R$. The domain covered by the contracting cosmological coordinates is shaded by light gray. The world line of a static observer $\kappa$ located in $R_{\rm out}<R<R_{\rm cos}$ is depicted by a bit thick curve.
The causal past $J^-(P)$ of the observer $\kappa$ at the event $P$ is the shaded domain also by light gray. The intersection of
the domain covered by the contracting cosmological coordinates and $J^-(P)$ is shaded by dark gray. There is a white hole in the causal past of the observer $\kappa$.
Dashed curves are spacelike hypersurfaces of $\t_-=$ constant.	}
\label{RNdSPen}
\end{figure}
%%%%%%%%%%%

We adopt the spherical polar coordinate system $(r,\theta, \phi)$ in which the particle is located at the origin, i.e., $r=r_1$. Then we have
\begin{align}
	& ds^2=-U^{-2}d \t_{\pm}^2+U^2
	\left( dr^2+r^2d\O^2 \right) \label{RNdSWc},
\end{align}
where
\begin{align}
	& U=\pm H\t_\pm+\f{M}{r} \label{RNdSW}.
\end{align}
The familiar form of the metric of the RNdS spacetime is given in the static and spherical polar coordinate system $(T,R,\theta,\phi)$ related with the cosmological
coordinates $(\t_\pm,r,\theta,\phi)$ through
\begin{align}
	& \pm H\t_{\pm}=\exp\left(\pm HT - \int \f{H^{2}R^2}{\left(R-M \right)V\left(R\right)}dR\right),\label{CoordTrtT} \\
	& \pm H\t_\pm r =R-M \quad .
	\label{cosmor}
\end{align}
Then the metric in the static coordinates is given as
\begin{align}
& ds^2=-V(R)dT^2+V^{-1}(R)dR^2+R^2d \O^2 \label{RNdSsc},
\end{align}
where
\begin{align}
& V(R)=\left(1-\f{M}{R}\right)^2-H^2R^2.
\end{align}
In the case of $1-4HM>0$, there are three positive real roots of $V\(R\)=0$:
\begin{align}
&R_{\text{in}}=-\f{1-\sqrt{1+4HM}}{2H}
\,, \label{Rin} \\
&R_{\text{out}}=\f{1-\sqrt{1-4HM}}{2H}
\,, \label{Rout}
\\
&R_{\text{cos}}=\f{1+\sqrt{1-4HM}}{2H}
\,. \label{Rcos}
\end{align}
$R_{\text{in}}$, $R_{\text{out}}$, and $R_{\text{cos}}$ are called the inner horizon, the outer horizon, and the cosmological horizon, respectively.
This solution describes the asymptotic de Sitter spacetime with a black hole if and only if $1-4HM>0$ holds, and in this case, the event horizon is located at $R=R_{\rm out}$.

The maximally extended conformal diagram of the extreme RNdS with $1-4HM>0$ is given in Fig.\ref{RNdSPen}.
The contracting cosmological coordinates $(\t_{-},r)$ cover a domain shaded by light gray. In this diagram, the world line of a static observer $\kappa$
located in $R_{\rm out}<R<R_{\rm cos}$ is depicted. The causal past $J^-(P)$ of an event $P$ on the world line of $\kappa$ is also depicted by the shaded domain also by light gray. The intersection of the domain covered by the cosmological coordinate and $J^-(P)$ is shaded by dark gray.
Note that there is a white hole in $J^-(P)$, and hence the observer $\kappa$ can take a picture of a
black hole shadow at the event $P$.

%%%%%%%%%%%%%%%%%%%%%%%%%%%%%%%%%%%%%%%%%%%%%%%%%%%%%%%%%%%%%%%%%%%%%%%%%
\section{Case of two particles}
\label{2-particles}
%%%%%%%%%%%%%%%%%%%%%%%%%%%%%%%%%%%%%%%%%%%%%%%%%%%%%%%%%%%%%%%%%%%%%%%%%

In this paper, our main interest is in the case of two particles with identical mass $M_1=M_2=M/2<1/8H$
in the contracting cosmological coordinates.
This solution describes the coalescence of two black holes to one black hole\cite{PhysRevD.49.840,PhysRevD.52.796,PhysRevD.58.121501}.
As for space coordinates, we adopt the cylindrical coordinates $(\r,\phi,z)$, since the spacetime has axisymmetry.
The metric function $U$ is given as
\begin{align}
U&=-H\t_- \nonumber \\
&+\frac{M}{2\sqrt{\r^2+(z-l)^2}}+\frac{M}{2\sqrt{\r^2+(z+l)^2}},
\end{align}
where $l$ is a positive constant. In order to see whether the black hole shadows merge with each other,
we study the global structure of the hypersurface specified by $z=0$ which extends through just the middle of two particles.
Hereafter this hypersurface is denoted by $\varSigma$. The intrinsic metric of $\varSigma$ is given in the form
\begin{align}
ds_\varSigma^2=-W^{-2}d\t_-^2+W^2\left(d\r^2+\r^2 d\phi^2\right), \label{intrinsic}
\end{align}
where
\begin{align}
W(\t_-,\r)=-H\t_-+\frac{M}{\sqrt{\r^2+l^2}}. \label{intrinsicW}
\end{align}
The hypersurface $\varSigma$ is totally geodesic, i.e., its extrinsic curvature vanishes.

%%%%%%%%%%%%%%%%%%%%%%%%%%%%%%%%%%%%%%%%%%%%%%%%%%%%%%%%%%%%%%%%%%%%%%%%%
\subsection{Asymptotic behavior}
\label{Asympt}
%%%%%%%%%%%%%%%%%%%%%%%%%%%%%%%%%%%%%%%%%%%%%%%%%%%%%%%%%%%%%%%%%%%%%%%%%

The purpose in this section is to draw the conformal diagram of the domain in $\varSigma$ covered by the coordinates $(\t_-,\r,\phi)$, i.e.,
\begin{align}
-\infty<\t_-<\frac{M}{H \sqrt{\r^2+l^2}}~~~{\rm and}~~~0\leq\r<\infty, \label{domain}
\end{align}
and $0\leq\phi<2\pi$.

From Eqs.~\eqref{intrinsic} and \eqref{intrinsicW},
the lower bound of $\r$, i.e., $\rho=0$ is a timelike curve along which $ds^2=-W^{-2}d\t_-^2<0$ with $W=-H\t_-+M/l$.

In the domain of $\r \gg l$, the metric function $W$ approximately behaves as the extreme RNdS spacetime in the manner,
\begin{align}
W= -H\t_-+\frac{M}{\r}\left[1 +{\cal O}\left(\frac{l^2}{\r^2}\right)\right].
\end{align}
Hence, $\varSigma$ asymptotically approaches to the extreme RNdS spacetime with mass $M$ in the limit of $\rho\rightarrow\infty$ with $\t_-$
negative and constant. This limit corresponds to an infinity which is a point denoted by $i^0$ in the conformal diagram.
The limit $(\t_-,\r) \rightarrow (0,\infty)$ is classified into two categories in accordance with the quantity
\begin{align}
R_{\rm lim}^+\equiv \lim_{\t\rightarrow0,\r\rightarrow\infty} \left(-H\t_- \r\right)+M.
\end{align}
The limit of $R^{+}_{\rm lim}=R_{\rm cos}$ is not an infinity but the cosmological horizon across which $C^2$ extension is possible
\cite{PhysRevD.49.840}. By contrast, the limit of $R_{\rm out}<R^+_{\rm lim}<R_{\rm cos}$ is an infinity which is a point denoted by $i^+$ in the conformal diagram.

In the domain of $-H\t_-\gg M/l$,
the metric function $W$ behaves as
\begin{align}
W= -H\t_-\left[1+{\cal O}\left(\frac{M}{lH\t_- }\right)\right].
\end{align}
Thus, the metric function asymptotically approaches to the de Sitter one in the limit of $\t_-\rightarrow-\infty$.
Also in this case, this limit is classified in accordance with the quantity
\begin{align}
R_{\rm lim}^-=\lim_{\t_-\rightarrow-\infty}\left(-H\t_- \rho\right)+M.
\end{align}
In the case that the limit is taken with $\r$ constant, $R_{\rm lim}^{-}$ positively diverges. This limit corresponds to an infinity
which extends over a spacelike direction as in the case of de Sitter spacetime.
By contrast, in the case that the limit is taken with $\r\rightarrow0$ so that $R_{\rm lim}^-$ is positive and finite or vanishes,
the limit is an infinity which is a point denoted by $i^-$ in the conformal diagram.

At the upper bound of $\t_-$ in Eq.~\eqref{domain}, i.e.,
\begin{align}
\t_-=\frac{M}{H\sqrt{\r^2+l^2}}, \label{singularity}
\end{align}
there is a scalar polynomial curvature singularity \cite{hawking_ellis_1973} at which $W$ vanishes.
Equation \eqref{singularity} implies that the singularity exists in $\t_->0$ and has endpoints at $(\t_-,\rho)=(M/Hl,0)$ and $(0,\infty)$.
In order to see the causal property of this singularity, we introduce a conformal metric defined as
\begin{equation}
d\tilde{s}^2=H^2W^2ds_\varSigma^2,
\end{equation}
and adopt $W$ as a coordinate instead of $\t_-$. Note that the causal structure of the (2+1)-dimensional spacetime with $d\tilde{s}^2$ is the same as that
with $ds_\varSigma^2$, since the null structure of both spacetimes are the same as each other. From Eq.~\eqref{intrinsicW}, we have
\begin{align}
d\t_-=-\frac{1}{H}dW-\frac{M\r}{H\left(\r^2+l^2\right)^{3/2}}d\r,
\end{align}
and substituting this equation into Eq.~\eqref{intrinsic}, we obtain
\begin{align}
d\tilde{s}^2=&-\left(dW+\frac{M\r}{\left(\r^2+l^2\right)^{3/2}}d\r\right)^2 \nonumber \\
&+H^2W^4\left(d\r^2+\r^2 d\phi^2\right).
\end{align}
The induced conformal metric on the singularity at which $W$ vanishes is then given by
\begin{align}
d\tilde{s}_{\rm sng}=-\frac{M^2\r^2 }{\left(\r^2+l^2\right)^3} d\rho^2 <0.
\end{align}
This equation implies that the singularity $W=0$ is timelike.
Note that the singularity $W=0$ is not an infinity since null geodesics can reach there at finite affine parameter. (See Appendix \ref{geodesics}).

To summarize, the boundary of the domain on $\varSigma$ covered by the coordinates $(\t_{-},\r,\phi)$ in the conformal diagram is
classified into the following seven categories :
\begin{enumerate}[ \textrm{(}i\textrm{)} ]
 \item $\r=0$: timelike coordinate boundary.
 \item $\r\rightarrow \infty$ with $\t_-$ negative and constant: an infinity which is a point denoted by $i^0$.
 \item $\r\rightarrow \infty$ and $\t_-\rightarrow0$ with $R_{\rm lim}^+=R_{\rm cos}$: the cosmological horizon which extends over a null direction as in the case of
 the RNdS spacetime.
 \item $\r\rightarrow \infty$ and $\t_-\rightarrow0$ with $R_{\rm out}<R_{\rm lim}^+<R_{\rm cos}$: an infinity which is a point denoted by $i^+$.
 \item $\t_{-}\rightarrow-\infty$ with $R_{\rm lim}^-=\infty$ : an infinity which extends over a spacelike direction as in the case of the de Sitter spacetime.
 \item $\t_{-}\rightarrow-\infty$ with $R_{\rm lim}^-$ finite: an infinity which is a point denoted by $i^-$.
 \item $W=0$: timelike scalar polynomial curvature singularity.
\end{enumerate}

%%%%%%%%%%%%%%%%%%%%%%%%%%%%%%%%%%%%%%%%%%%%%%%%%%%%%%%%%%%%%%%%%%%%%%%%%
\subsection{Event horizon}
\label{event_horizon}
%%%%%%%%%%%%%%%%%%%%%%%%%%%%%%%%%%%%%%%%%%%%%%%%%%%%%%%%%%%%%%%%%%%%%%%%

In order to draw the event horizon in the conformal diagram of $\varSigma$, we consider the intersection
between the event horizon and $\varSigma$, which is a circle with temporally varying radius $\r=\r(\t_-)$. Since the spacetime has
reflection symmetry with respect to $\varSigma$ and furthermore $\varSigma$ is totally geodesic,
the intersection between the event horizon and $\varSigma$ is generated by null geodesics.

The event horizon of the KT spacetime with two particles was studied by one of the present authors and his collaborators \cite{PhysRevD.58.121501}.
The numerical result given in this paper implies that the end point of the null geodesic generators of the event horizon on $\varSigma$ is located at $\r=0$
and at finite negative $\t_-$. This result can be verified analytically as follows.

The event horizon of the RNdS spacetime with mass $M$ in the contracting cosmological coordinate is located on
\begin{align}
r=\frac{R_{\rm out}-M}{-H\t_-}. \label{RNdS-EH}
\end{align}
We can easily verify that this is a solution of the future-directed outgoing radial null condition
\begin{equation}
\frac{d\t_-}{dr}=\left(-H\t_-+\frac{M}{r}\right)^2. \label{RNdS-NullCd}
\end{equation}
By contrast, the outgoing radial null condition on $\varSigma$ is
\begin{align}
\frac{d\t_-}{d\r}=\left(-H\t_-+\frac{M}{\sqrt{\r^2+l^2}}\right)^2. \label{KT-NullCd}
\end{align}
The null geodesic generators of the event horizon on $\varSigma$ satisfy Eq.~\eqref{KT-NullCd}. Since, as shown in Sec.~\ref{Asympt},
$\varSigma$ approaches to the RNdS in the limit of $\r\rightarrow\infty$, the null geodesic generators behave as Eq.~\eqref{RNdS-EH};
\begin{align}
\r \longrightarrow \frac{R_{\rm out}-M}{-H\t_-}~~~{\rm for}~~~\t_-\rightarrow 0-. \label{EH_asympt}
\end{align}
The null geodesic generators of the event horizon are then in the domain $\t_-<0$. If the null geodesic generators intersect $\r=0$, then it is the endpoint
of them by the axisymmetry.

Here note that
\begin{align}
\frac{d\t_-}{d\r}<\left(-H\t_-+\frac{M}{\r}\right)^2 \label{EH-ineq}
\end{align}
holds for the null geodesic generators.
Since the asymptotic solution \eqref{EH_asympt} exactly satisfies the equation obtained by replacing the sign of inequality by an equal sign
in Eq.~\eqref{EH-ineq}, Eq.~\eqref{EH-ineq} implies that the null geodesic generators should satisfy
\begin{align}
\t_->\frac{R_{\rm out}-M}{-H\r}
~~{\rm or}~{\rm equivalently}~~\r<\frac{R_{\rm out}-M}{-H\t_-}. \label{EH-ineq-0}
\end{align}
Note that this inequality does not imply the existence of a lower bound on $\t_-$ at $\r=0$, and hence we need further consideration.

The following inequality also holds on the null geodesic generators:
\begin{align}
\frac{d\t_-}{d\r}<\left(-H\t_-+\frac{M}{l}\right)^2. \label{EH-ineq-1}
\end{align}
Let $(\t_-,\r)=(\t_{\rm e},\r_{\rm e})$ denote an event on a null geodesic generator. Here note that $\t_{\rm e}$ is negative, whereas $\r_{\rm e}$ is positive.
Then, it is easy to obtain a solution of the differential equation
obtained by replacing the sign of inequality by an equal sign in Eq.~\eqref{EH-ineq-1}, which intersects
a null geodesic generator at $(\t_-,\r)=(\t_{\rm e},\r_{\rm e})$:
\begin{align}
\t_-=\frac{M}{H l }-\frac{1}{H^2}\left[\r-\r_{\rm e}+\frac{1}{H\left(-H\t_{\rm e}+\dfrac{M}{l}\right)}\right]^{-1}.
\end{align}
Equation \eqref{EH-ineq-1} implies that the null geodesic generators satisfy, for $\r<\r_{\rm e}$,
\begin{align}
\t_- >\frac{M}{H l }-\frac{1}{H^2}\left[\r-\r_{\rm e}+\frac{1}{H\left(-H\t_{\rm e}+\dfrac{M}{l}\right)}\right]^{-1}. \label{EH-ineq-2}
\end{align}
If the following inequality
\begin{align}
\r_{\rm e}-\frac{1}{H\left(-H\t_{\rm e}+\dfrac{M}{l}\right)} < 0 \label{crucial-ineq}
\end{align}
holds, Eq.~\eqref{EH-ineq-2} gives a lower bound of $\t_-$ of null geodesic generators on $\varSigma$ at $\r=0$ as
\begin{align}
\t_- \Big|_{\r=0}
&>\frac{M}{H l } \left[\r_{\rm e}-\frac{1}{H\left(-H\t_{\rm e}+\dfrac{M}{l}\right)}\right]^{-1} \nonumber \\
&\times \left[\r_{\rm e}+\frac{l}{HM}-\frac{1}{H\left(-H\t_{\rm e}+\dfrac{M}{l}\right)}\right].
\end{align}
We can easily see that if Eq.~\eqref{EH-ineq-2} is satisfied, the right-hand side of this inequality is negative and finite.

A remaining task is to show that there is an event $(\t_{\rm e},\r_{\rm e})$ that satisfies Eq.~\eqref{crucial-ineq}.
We consider following two curves in the spacetime diagram $(\t_-,\r)$;
\begin{align}
\rho&=\frac{1}{H\left(-H\t_-+\dfrac{M}{l}\right)},  \label{curve-1}\\
\rho&=\frac{R_{\rm out}-M}{-H\t_-}. \label{curve-2}
\end{align}
An intersection of these two curves is easily obtained as
\begin{align}
\t_-&=\t_{\rm i}\equiv -\frac{HMR_{\rm out}^2}{l \left(1-H^2R_{\rm out}^2\right)}, \\
\r&=\r_{\rm i} \equiv \frac{l \left(1-H^2R_{\rm out}^2\right)}{HM}.
\end{align}
It is not difficult to see $1-H^2R_{\rm out}^2>0$, and hence two curves \eqref{curve-1} and \eqref{curve-2} intersect with each other in the domain
of $\t_-<0$ and $\r>0$. We can see that the following equations hold;
\begin{align}
&\frac{d}{d\t_-}\left.\frac{1}{H\left(-H\t_-+\dfrac{M}{l}\right)}\right|_{\tau_-=\t_{\rm i}}=\left[\frac{ l \left(1-H^2R_{\rm out}^2\right)}{M}\right]^2, \\
&\frac{d}{d\t_-}\left.\frac{R_{\rm out}-M}{-H\t_-}\right|_{\tau_-=\t_{\rm i}}=\left[\frac{l \left(1-H^2R_{\rm out}^2\right)}{HMR_{\rm out}}\right]^2
\end{align}
Because of $HR_{\rm out}<1/2$, we have
\begin{align}
\frac{d}{d\t_-}\left.\frac{R_{\rm out}-M}{-H\t_-}\right|_{\tau_-
=\t_{\rm i}}>\frac{d}{d\t_-}\left.\frac{1}{H\left(-H\t_-+\dfrac{M}{l}\right)}\right|_{\tau_-=\t_{\rm i}},
\end{align}
and hence we obtain
\begin{align}
\frac{R_{\rm out}-M}{-H\t_-}<\frac{1}{H\left(-H\t_-+\dfrac{M}{l}\right)}
\end{align}
for $\t<\t_{\rm i}$ or equivalently $\r<\r_{\rm i}$. By virtue of Eq.~\eqref{EH-ineq-0}, Eq.~\eqref{crucial-ineq} holds if $\t_{\rm e}<\t_{\rm i}$ holds.
This means that there is an endpoint of null geodesic generators on $\varSigma$.

%%%%%%%%%%%%%%%%%%%%%%%%%%%%%%%%%%%%%%%%%%%%%%%%%%%%%%%%%%%%%%%%%%%%%%%%%
\subsection{Conformal diagram}
\label{conformal_diagram}
%%%%%%%%%%%%%%%%%%%%%%%%%%%%%%%%%%%%%%%%%%%%%%%%%%%%%%%%%%%%%%%%%%%%%%%%

%%%%%%%%%%%
\begin{figure}[htbp]
	\includegraphics[width=8cm,height=8cm,keepaspectratio]{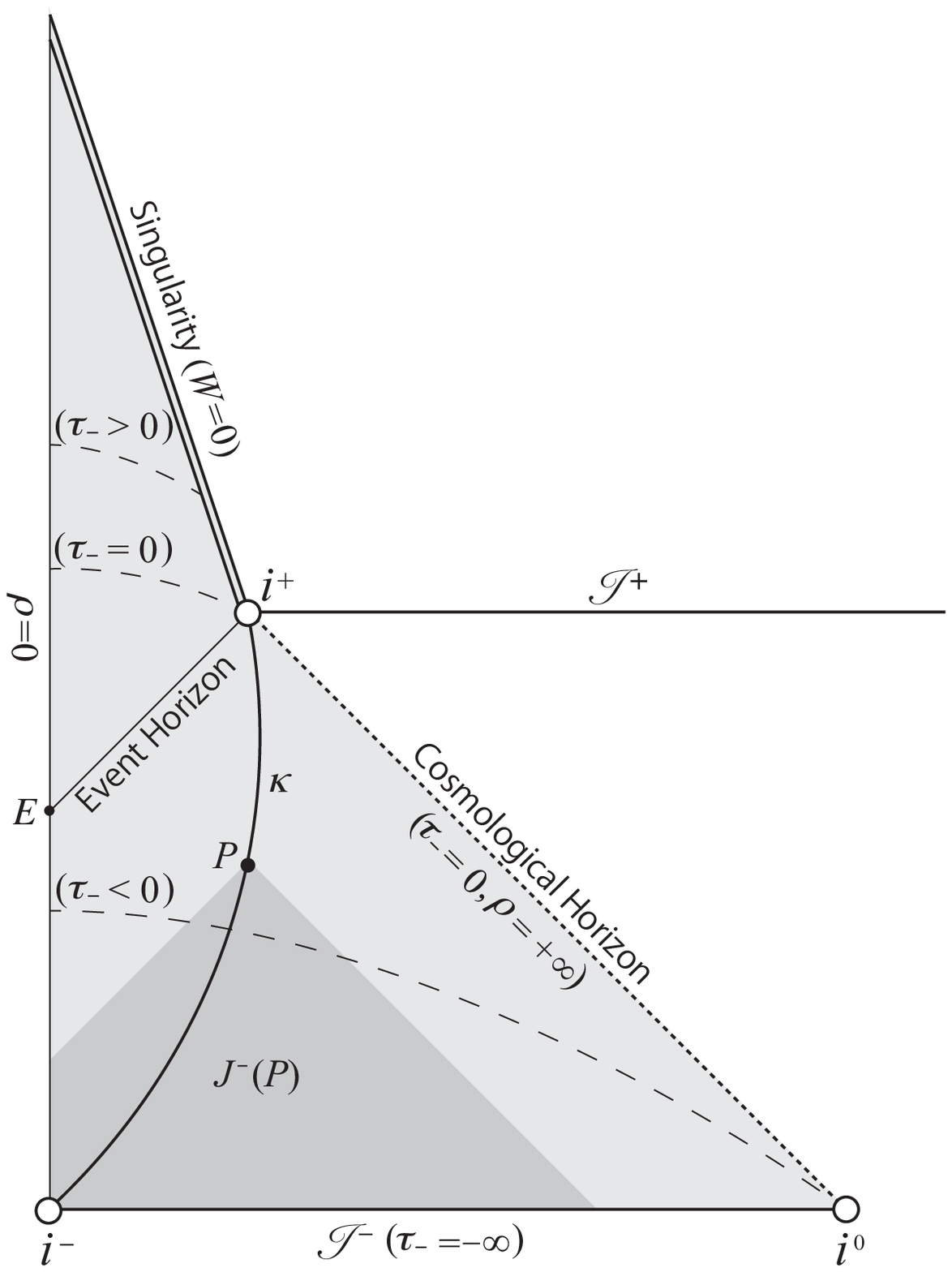}
\caption{
The conformal diagram of $\varSigma$ is depicted. The domain covered by the contracting cosmological coordinate $(\t_-,\r,\phi)$
is shaded by light gray. The event horizon is represented by a thin line and the end point of its null geodesic generators is the event $E$.
The world line of an observer $\kappa$ who keeps the distance from the middle of two black holes constant
is depicted by a bit thick curve.
The intersection of the domain covered by the contracting cosmological coordinates and
the causal past of the observer $\kappa$ at the event $P$ is shaded by dark gray.
Dashed curves are spacelike hypersurfaces of $\t_-=$ constant.
There is no white hole in intersection of $\varSigma$ and the causal past of any observer outside the black hole.
}
	\label{Concl}
\end{figure}
%%%%%%%%%%%

The conformal diagram of $\varSigma$ is depicted in Fig. \ref{Concl}. The domain covered by the contracting cosmological coordinates $(\t_-,\r,\phi)$
is shaded by light gray.
Each point except on the boundary in this diagram corresponds to a circle $0\leq \phi<2\pi$. The event horizon is represented by a thin line and the end point
of its null geodesic generators is the event $E$.
The world line of an observer $\kappa$ who keeps the distance from the middle of two black holes constant is depicted as a bit thick curve.
The intersection of the domain covered by the contracting cosmological coordinates and
the causal past of the observer $\kappa$ at the event $P$ is shaded by dark gray, and there is no white hole in it.
Since a black hole shadow is an image of a white hole, there is no null geodesic which makes black hole shadow on $\varSigma$.
Here again note that $\varSigma$ is the hypersurface going through just the middle of the two black holes.
This fact implies that black hole shadows taken by observers on $\varSigma$ do not merge
with each other in the KT spacetime with two identical black holes.

%%%%%%%%%%%%%%%%%%%%%%%%%%%%%%%%%%%%%%%%%%%%%%%%%%%%%%%%%%%%%%%%%%%%%%%%%
\section{CONCLUDING REMARKS}
\label{concluding_remarks}
%%%%%%%%%%%%%%%%%%%%%%%%%%%%%%%%%%%%%%%%%%%%%%%%%%%%%%%%%%%%%%%%%%%%%%%%
By analyzing the null geodesic generators of the event horizon in the KT spacetime with two particles, we obtain the conformal diagram of the hypersurface $\varSigma$
which passes just in the middle of the two particles with identical mass in Sec.\ref{2-particles}.
We showed analytically that there is the endpoint of null geodesic generators of the event horizon on $\varSigma$
and there is no intersection between a white hole and $\varSigma$.
These facts imply that any observer restricted on $\varSigma$
can never see the merger of black hole shadows.

Here it is worthwhile to notice that the number of black holes in their merger process is coordinate dependent notion\cite{PhysRevD.58.121501};
we can adopt a time slicing in which three black holes merge into two and eventually into one black hole even in the case of the KT spacetime with two particles
investigated in Sec.~\ref{2-particles}. By contrast, the number of black hole shadows is observable and thus should not depend on the choice of coordinates.
In addition, it might be conserved for any observers. However, in order to show that this conjecture is true, we need to investigate whether black hole shadows
taken by any observers do not merge. This issue is out of the scope of the present paper and future subject.

As shown by Yumoto, et al \cite{Yumoto_2012}, an interval between two black hole shadows becomes indefinitely narrower as time elapses, and hence
hence those will eventually look like one merged black hole shadow due to the limitation of the observational sensitivity.
Furthermore, the redshift effects on the photons
coming through the space between the black hole shadows become larger as time elapsed, or equivalently,
as the shadows becomes closer to each other. (See Appendix B).

%\newpage
%\section*{Acknowledgments}
\acknowledgments
We would like to thank Hirotaka Yoshino, Shoichiro Miyashita and Chul-Moon Yoo for giving us useful information and making crucial comments on our study.

%%%%%%%%%%%%%%%%%%%%%%%%%
%%%%%%%%%%%%%%%%%%%%%%%%%
\appendix

%%%%%%%%%%%%%%%%%%%%%%%%%%%%%%%%%%
\section{What is black hole shadow?}
\label{BH-shadow}
%%%%%%%%%%%%%%%%%%%%%%%%%%%%%%%%%%

Some confusion about the black hole shadow might exist. In order to avoid it, we reconsider what is the black hole shadow here.
In accordance with Gralla, Holz and Wald\cite{PhysRevD.100.024018},
we consider the case that the black holes are illuminated by a distant, uniform, isotropically emitting spherical screen surrounding both of an observer and the black holes.
In this situation, the observer will find dark domains on the celestial sphere, which are called black hole shadows. In the case of the Kerr spacetime,
the shape of the black hole shadow was given by Bardeen\cite{1973blho.conf..215B}.

\begin{figure}[htbp]
	\includegraphics[width=8cm,height=8cm,keepaspectratio]
	{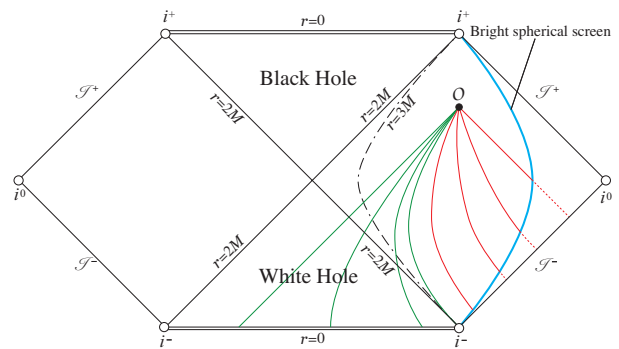}
	\caption{Null geodesics arriving at the event $O$ outside the black hole are depicted in the conformal diagram of the Schwarzschild spacetime. The bright spherical screen
	is represented by a thick blue curve. Red curves are null geodesics which intersect with the bright spherical screen, whereas green curves are null geodesics
	which do not intersect with the bright spherical screen. Hence, the direction cosines of null geodesics represented by red curves are in a bright domain
	on the celestial sphere of the observer, whereas those represented by the green curves are in a dark domain.
	}
	\label{BH-shadow-1}
\end{figure}

\begin{figure}[htbp]
	\includegraphics[width=8cm,height=8cm,keepaspectratio]
	{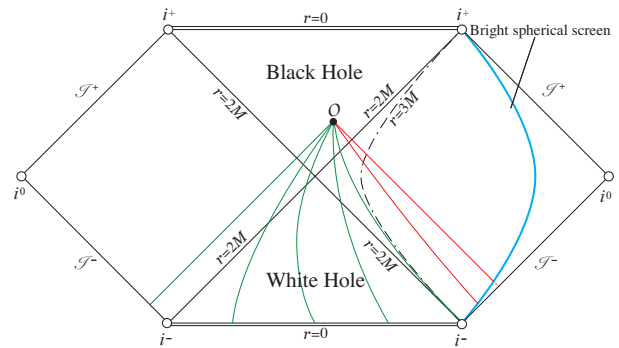}
	\caption{The same as Fig.~\ref{BH-shadow-1} but the event $O$ is inside the black hole. Even in this case, the observer see a black hole shadow at the event $O$. 	}
	\label{BH-shadow-2}
\end{figure}

For simplicity, first we consider the case that an observer detects photons at the event $O$ in the Schwarzschild spacetime with mass $M$.
The bright domain on the celestial sphere is generated by the direction cosines of photons moving along null geodesics
from the bright spherical screen to the event $O$.  By contrast, the dark domain on the celestial sphere, i.e., black hole shadow is generated by direction cosines of null geodesics that do not intersect with the bright spherical screen in the causal past of $O$, which is usually denoted by $J^-(O)$.  Figures \ref{BH-shadow-1} and \ref{BH-shadow-2} depict the situations explained above.   Figure \ref{BH-shadow-1} depicts the case that the event $O$ is located outside the black hole, whereas Fig.~\ref{BH-shadow-2} shows the same but the event $O$ is inside the black hole. In these figures, the world lines of the bright spherical screen is represented by a thick blue curve. The null geodesics represented by green curves do not intersect with the bright spherical screen in $J^-(O)$, and hence generate the black hole shadow. By contrast, the null geodesics represented by red curves generate the bright domain on the celestial sphere since they intersect with the bright spherical screen in $J^-(O)$. We can see from Fig.~\ref{BH-shadow-2} that the observer falling from the right-hand side asymptotically flat domain with the bright spherical screen can take a picture of the black hole shadow even after entering the black hole\cite{chang2019black}.
It is not so difficult to see, by investigating null geodesics with the critical impact parameter $3\sqrt{3}M$, that the angular radius of the black hole shadow seen by a marginally bound freely falling observer is equal to ${\rm arctan}(12\sqrt{3}/23)\simeq 0.23\pi$ at the moment when the observer arrives at the event horizon\cite{Yoshino_Fujioka_Nakao}.
This fact {\rm definitely} implies that the black hole shadow does not come from the absorption of photons by the black hole.

In order to get the black hole shadow theoretically, the so-called ray tracing method is efficient; we trace the null geodesics from the event $O$ in the past direction and investigate whether they intersect with the bright spherical screen. Usually, we assume that the observer like us is located outside the black hole as in Fig.~\ref{BH-shadow-1}.
In this case, in the ray-tracing method, we stop tracing a null geodesic in the past direction from the event $O$ and regard its direction cosine as an element of the black hole shadow
if they reach a sphere of $r=2M$. As can be seen from Fig.~\ref{BH-shadow-1}, the sphere of $r=2M$ is not the event horizon but the white hole horizon. Thus,
we may say that the black hole shadow is an image of a white hole horizon\cite{PhysRevD.99.044027}.

%%%%%%%%%%%
%%%%%%%%%%%

%%%%%%%%%%%
%%%%%%%%%%%

\begin{figure}[htbp]
	\includegraphics[width=8cm,height=8cm,keepaspectratio]
	{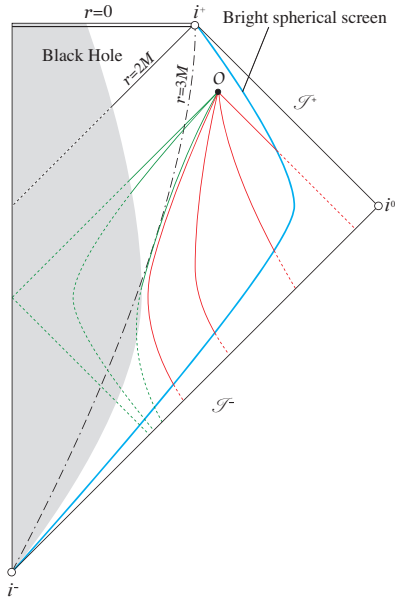}
	\caption{The conformal diagram of the spacetime in which a spherically symmetric black hole forms by the gravitational collapse of matter represented by
	a grayed domain.  All null geodesics arriving at the event $O$ intersect with the bright spherical screen. The null geodesics represented by green curves
	go through the collapsing object, whereas those represented by red curves do not. Even in this case,
	the observer can see a dark image similar to that observed in the Schwarzschild spacetime, since the green null geodesics suffers the kinematical and gravitational redshift.}
	\label{BH-shadow-3}
\end{figure}

Black holes in our Universe will form through gravitational collapse of massive objects,
and hence there will be no white hole horizon. However, even in these cases, the dark images will appear.
We consider the case that a black hole forms by the gravitational collapse of matter in spherically symmetric asymptotically flat spacetime as depicted in Fig.~\ref{BH-shadow-3}.
We can see from this figure that all null geodesics passing through the event $O$ intersect with the bright spherical screen in $J^-(O)$. Here we should note the fact that  there are null geodesics represented by green curves which hit the collapsing object after they leave the bright spherical screen. If the collapsing object is not transparent to the photons, the observer detects no photon moving along such null geodesics from the bright spherical screen. Hence the direction cosines of such null geodesics generate a dark image on the celestial sphere, if the collapsing object emits nothing. Even if the collapsing object emit photons with finite energy, those photons suffer strong kinematical and gravitational redshift and hence cannot be detected due to the limitation of the detectability at sufficiently late stage of the gravitational collapse. As a result, a dark image will eventually appear, even if the collapsing object emits radiation\cite{PhysRevD.100.084062}.  If the collapsing object is transparent to photons, what happens?  Also, in this case, a dark image will appear, since photons going through the collapsing object suffer very strong redshift due to a kind of the so-called Rees-Sciama effects\cite{rees_sciama} at the late stage of the gravitational collapse (see, for example, Appendix A of Ref.~\cite{PhysRevD.99.044027}).
The shape of the dark image is also determined
by null geodesics with the critical impact parameter $3\sqrt{3}M$. Also in this case, even if the event $O$ is inside the black hole, the dark image will appear \cite{Yoshino_Fujioka_Nakao}.  In our Universe, the black hole shadow will be a silhouette of a collapsing object.

%%%%%%%%%%%
%%%%%%%%%%%

A black hole shadow is not a silhouette of a black hole.

%%%%%%%%%%%%%%%%%%%%%%%%%%%%%%%%%%
\section{Null geodesics to the singularity $W=0$ in the case of two particles of the KT spacetime}
\label{geodesics}
%%%%%%%%%%%%%%%%%%%%%%%%%%%%%%%%%%

The proper time $t_{\rm sng}(\r)$ from $\t_-=0$ to the singularity along a timelike curve of $\r=$ constant is
\begin{align}
t_{\rm sng}(\r)&=\int_0^{\t_{\rm sng}}\left(-H\t_-+\frac{M}{\sqrt{\r^2+l^2}}\right)^{-1}d\t_- \nonumber \\
&=+\infty,
\end{align}
where
\begin{align}
\t_{\rm sng}=\frac{M}{H\sqrt{\r^2+l^2}}.
\end{align}
Thus, the singularity seems to be located at infinity. However, it is not true.
In order to see this fact, we consider the null geodesics on $\varSigma$ and along $\r=0$ normal to $\varSigma$.
Since $\varSigma$ is totally geodesic, the geodesics on $\varSigma$ are also geodesics in the spacetime.

The Lagrangian of a geodesic is given as
\begin{align}
	\mathcal{L}=-\frac{1}{U^2}\left(\frac{d\t}{d\lambda}\right)^2+U^2\left[\left(\frac{d\r}{d\lambda}\right)^2+\r^2\left(\frac{d\phi}{d\lambda}\right)^2+\left(\frac{dz}{d\lambda}\right)^2 \right],
	\notag
\end{align}
where $\lambda$ is the affine parameter. The variation of $\phi$ leads to
\begin{align}
\frac{d}{d\lambda}\left(U^2\r^2\frac{d\phi}{d\lambda}\right)=0.
\end{align}
Then we have
\begin{align}
\frac{d\phi}{d\lambda}=\frac{L}{U^2\r^2}, \label{AM}
\end{align}
where $L$ is an integration constant which corresponds to the angular momentum. Hereafter, we focus on the case of $L=0$.

\begin{figure}[htbp]
	\includegraphics[width=8cm,height=8cm,keepaspectratio]
	{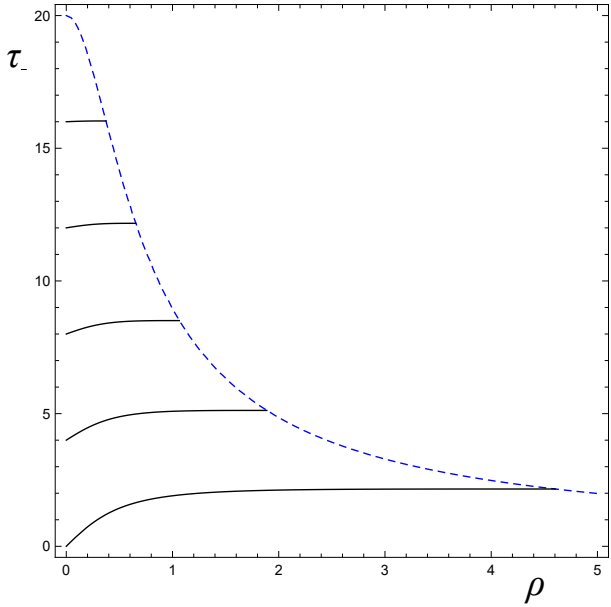}
	\caption{
	On $\varSigma$, null geodesics with $L=0$ inside the event horizon are depicted as solid lines while the singularity $W=0$ is represented by a dashed  curve. We set $M=1$, $H=10^{-1}$, $l=1/2$, and choose $(\tau_-, \rho)=(16,0), (12,0), (8,0), (4,0), (0,0)$ as an initial condition in order from the top curve. Each null geodesic reaches the singularity with finite affine length. Hence, the singularity except at $(\t_-,\r)=(M/Hl,0)$ on $\varSigma$ is actually not infinity.
	}
	\label{nullFig}
\end{figure}

First we consider null geodesics on $\varSigma$. The geodesic equations are given as
\begin{align}
&\frac{d^2\t_-}{d\lambda^2}-\frac{\partial \ln W^2}{\partial \r}\frac{d\r}{d\lambda} \frac{d\t_-}{d\lambda}=0, \label{g-eq-t}\\
&\frac{d^2\r}{d\lambda^2}+\frac{\partial \ln W^2}{\partial \t_-}\frac{d\t_-}{d\lambda}\frac{d\r}{d\lambda} =0, \label{g-eq-r}
\end{align}
where we have used the null condition $\mathcal{L}=0$.
Since we consider null geodesics which hit singularity, the null condition is given as
\begin{align}
\frac{d\t_-}{d\lambda}=W^2\frac{d\r}{d\lambda}.
\end{align}
Then using this null condition, we rewrite Eq.~\eqref{g-eq-r} in the form
\begin{align}
\frac{d^2\lambda}{d\r^2}+2HW\frac{d\lambda}{d\r}=0. \label{g-eq-lam-S}
\end{align}
From the null condition, we have
\begin{align}
\frac{d\t_-}{d\r}=W^2.  \label{t-eq-S}
\end{align}
We numerically solve Eqs.~\eqref{g-eq-lam-S} and \eqref{t-eq-S} and depict
the result in Fig.\ref{nullFig}. The numerical results imply that there are null geodesics that reach the singularity $W=0$ except at
$(\t_-,\r)=(M/H l ,0)$ on $\varSigma$ with finite affine length.

%%%%%%%%%%%
%%%%%%%%%%%

In order to see whether $(\t_-,\r)=(M/H l,0)$ on $\varSigma$, or equivalently,  $(\t_-,\r,z)=(M/H l,0,0)$ is infinity, we study
null geodesics along $\r=0$ from a point in $-l<z<0$ to $z=0$. The geodesic equation is given as
\begin{align}
&\frac{d^2\t_-}{d\lambda^2}-\frac{\partial \ln X^2}{\partial z}\frac{dz}{d\lambda} \frac{d\t_-}{d\lambda}=0, \\
&\frac{d^2z}{d\lambda^2}+\frac{\partial \ln X^2}{\partial \t_-}\frac{d\t_-}{d\lambda}\frac{dz}{d\lambda} =0,\label{g-eq-z}
\end{align}
where we have used the null condition ${\cal L}=0$, and
\begin{align}
X=-H\t_- + \frac{M}{2(l-z)}+\frac{M}{2(l+z)}.
\end{align}
The null condition becomes
\begin{align}
\frac{d\t_-}{d\lambda}=X^2\frac{dz}{d\lambda}.
\end{align}
Using this equation, we rewrite Eq.~\eqref{g-eq-z} in the form
\begin{align}
\frac{d^2\lambda}{dz^2}+2HX\frac{d\lambda}{dz}=0. \label{g-eq-lam}
\end{align}
From the null condition, we have
\begin{align}
\frac{d\t_-}{dz}=X^2.  \label{t-eq}
\end{align}
Since we are interested in the null geodesic which hits $z=0$ at $\t_-=M/Hd$, we assume
\begin{align}
\t_-=\frac{M}{H l}+\sum_{n=1}^\infty\t_{(n)}z^n.
\end{align}
Then we have the left-hand side of Eq.~\eqref{t-eq} as
\begin{align}
\frac{d\t_-}{dz}=\sum_{n=0}^{\infty} (n+1)\t_{(n+1)}z^n.
\end{align}
By contrast, the right-hand side is rewritten in the form
\begin{align}
X^2&=\left[-\frac{M}{l}-H\sum_{n=1}^\infty\t_{(n)}z^n+\frac{M}{l}\sum_{n=0}^\infty\left(\frac{z}{l}\right)^{2n}\right]^2 \nonumber \\
&=\left[\sum_{n=1}^\infty\left(-H\t_{(n)}z^n + M l^{-(2n+1)}z^{2n}\right)\right]^2.
\end{align}
Then the solution for $|z|\ll l$ is written in the form
\begin{align}
\t_-&=\frac{M}{H l}+\frac{M^2}{5 l}\left(\frac{z}{l}\right)^5\biggl[1+\frac{10}{7}\left(\frac{z}{l}\right)^2 -\frac{HM}{4}  \left(\frac{z}{l}\right)^3 \nonumber \\
&+{\cal O}\left(\left(\frac{z}{l}\right)^4\right)\biggr].
\end{align}
Substituting this result into Eq.~\eqref{g-eq-lam} and integrating once, we obtain
\begin{align}
\frac{d\lambda}{dz} &=\exp\left(-2H \int^z X dz\right) \nonumber \\
&=C\exp\left[-\frac{2HM}{3}\left(\frac{z}{l}\right)^3+{\cal O}\left(\left(\frac{z}{l}\right)^5\right)\right] \nonumber \\
&=C\left[1-\frac{2HM}{3}\left(\frac{z}{l}\right)^3+{\cal O}\left(\left(\frac{z}{l}\right)^5\right)\right],
\end{align}
where $C$ is an integration constant. By integrating this equation, we have
\begin{align}
\lambda = Cz\left[1-\frac{HM}{6}\left(\frac{z}{l}\right)^3+{\cal O}\left(\left(\frac{z}{l}\right)^5\right)\right]+{\rm const.}
\end{align}
This result implies that there is a null geodesic that reaches the event $(\t_-,\r,z)=(M/Hl,0,0)$ on the singularity with finite affine length.

The singularity $W=0$ on $\varSigma$ is not infinity.

%%%%%%%%%%%%%%%%%%%%%%%%%%%%%%%%%%
\section{Redshift}
\label{redshift}
%%%%%%%%%%%%%%%%%%%%%%%%%%%%%%%%%%

Here, we numerically verify that a photon propagating through the neighborhood of the event horizon on the hypersurface $\varSigma$
is strongly gravitationally redshifted in the KT spacetime. The redshift is also caused by the kinematical effect due to the motion
of the emitter and the detector of the photon. Hence, in order to see the gravitational redshift, we usually assume that both the emitter and the detector are at rest.
Such an assumption is possible only when the spacetime is static or stationary. However, the KT spacetime
with two particles is neither static nor stationary. In order to estimate gravitational contribution to redshift,
we need to appropriately introduce an emitter and a detector which are approximately at rest. This is possible if $R_{\rm out}$ is much less than $R_{\rm cos}$, or equivalently,
$HM\ll1$. The domain of $R_{\rm out} \ll -H\t_- \rho +M < R_{\rm cos}$ is well approximated by
the RNdS spacetime, and hence we may define an emitter and a detector which are approximately at rest in this almost RNdS domain.

%%%%%%%%%%%%%%%%%%%%%%%%%
%%%%%%%%%%%%%%%%%%%%%%%%%
\begin{figure}[H]
	\includegraphics[width=8cm,height=8cm,keepaspectratio]
	{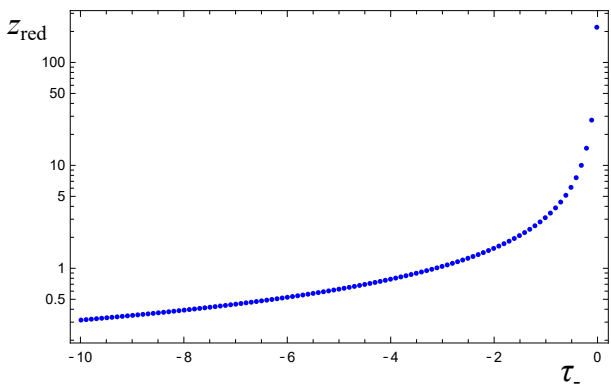}
	\caption{
	The redshift $z_{\rm red}$ suffered by a photon is depicted as a function of $\t_-$ at which the detector receives it. Both the detector and the emitter
	of the photon are approximately static. The photons are emitted for the time interval, $-13.16 \le \t_{-} \le -3.16$.
	The redshift suffered by photons going through the space between black hole shadows becomes indefinitely stronger as time elapses.}
	\label{RSfig}
\end{figure}

%%%%%%%%%%%%%%%%%%%%%%%%%
%%%%%%%%%%%%%%%%%%%%%%%%%

The radial coordinates of the detector $\r_{\rm d}$ and the emitter of photons $\r_{\rm e}$ are respectively given by
\begin{align}
	\r_{\text{d}}=\f{R_{\rm d}-M}{-H\t_-}, \quad
	\r_{\text{e}}=-\f{R_{\rm e}-M}{-H\t_-}, \notag
\end{align}
with $R_{\rm d}$ and $R_{\rm e}$ constant, where we assume that both $R_{\rm d}$ and $R_{\rm e}$ are much larger than $R_{\rm out}$ and less than $R_{\rm cos}$.
The four velocities of the detector $u_{\rm d}^\mu$ and the emitter $u_{\rm e}^\mu$ are
\begin{align}
u_{\rm d}^\mu&=N_{\rm d}\left(-\t_-,\r_{\rm d},0,0\right),  \\
u_{\rm e}^\mu&=N_{\rm e}\left(-\t_-,\r_{\rm e},0,0\right),
\end{align}
where
\begin{align}
N_{\rm d}&=\frac{W(\t_-,\r_{\rm d})}{\sqrt{\t_-^2-W^4(\t_-,\r_{\rm d})\r_{\rm d}^2}}, \\
N_{\rm e}&=\frac{W(\t_-,\r_{\rm e})}{\sqrt{\t_-^2-W^4(\t_-,\r_{\rm e})\r_{\rm d}^2}}.
\end{align}

We focus on the photons moving along null geodesics through $\r=0$, i.e., with $L=0$.
In order to find the world lines of photons, we numerically solve Eqs.~\eqref{g-eq-t} and \eqref{g-eq-r} from the emitter $\r=\r_{\rm e}$
to the detector $\r=\r_{\rm d}$. The null condition is imposed on the  initial conditions.
The tangent vector $k^\mu$ of the null geodesic is then
\begin{align}
k^\mu=\left(\frac{d\t_-}{d\lambda},\frac{d\r}{d\lambda},0,0\right).
\end{align}
The angular frequency of a photon is estimated as $\omega_{\rm e} \equiv -k_\mu u^\mu_{\rm e}$ at the emitter and $\omega_{\rm d} \equiv-k_\mu u^\mu_{\rm d}$
at the detector. Then the redshift $z_{{\rm red}}$ is defined as
%%%%%%%%%%%
\begin{equation}
1+z_{{\rm red}}=\frac{\omega_{\rm e}}{\omega_{\rm d}}.
\end{equation}
%%%%%%%%%%%

We show the numerical results for the case of $M=1$, $H=10^{-5}$, and $l=1$.
In this case, we have $R_{\text{out}} \approx 1$ and $R_{\text{cos}} \approx 10^5$, respectively, and set
both $R_{\rm d}$ and $R_{\rm e}$ to be $10^2$.
In Fig.~\ref{RSfig}, $z_{\rm red}$ is depicted as a function of $\t_-$ at which the detector receives the photon.
The photons emanate from the emitter for the time interval, $-13.16 \le \t_{-} \le -3.16$.
We can see from this figure that the gravitational redshift of the photon going through $\r=0$
becomes indefinitely stronger as time elapses. This is similar to the Rees-Sciama effect\cite{rees_sciama},
since the ``gravitational potential" in the neighborhood of $\r=0$ depends on time.

%%%%%%%%%%%%%%%%%%%%%%%%%
%%%%%%%%%%%%%%%%%%%%%%%%%
%\newpage
%%%%%%%%%%%%%%%%%%%%%%%%%
%%%%%%%%%%%%%%%%%%%%%%%%%
%%%%%%%%%%%%%%%%%%%%%%%%%
%%%%%%%%%%%%%%%%%%%%%%%%%
\bibliography{KT_BH_merger_ref}
%\bibliographystyle{JHEP}

%~~\\

%%%%%%%%%%%%%%%%%%%%%%%%%%
%%%%%%%%%%%%%%%%%%%%%%%%%%
%%%%%%%%%%%%%%%%%%%%%%%%%%
%%%%%%%%%%%%%%%%%%%%%%%%%%
%%%%%%%%%%%%%%%%%%%%%%%%%%
%%%%%%%%%%%%%%%%%%%%%%%%%%
%%%%%%%%%%%%%%%%%%%%%%%%%%
%%%%%%%%%%%%%%%%%%%%%%%%%%
\end{document}